\documentclass{llncs}
\usepackage[english]{babel} 
\usepackage{amsmath,amssymb,amsfonts,mathrsfs}
\usepackage[T1]{fontenc}
\usepackage[utf8]{inputenc}
\usepackage{dsfont}
\usepackage{booktabs}
\usepackage{algorithmic}
\usepackage{algorithm}
\usepackage[hidelinks]{hyperref}

\spnewtheorem{assumption}{Assumption}{\bfseries}{\itshape}
\spnewtheorem*{lemma*}{Lemma}{\normalshape\bfseries}{\itshape}
\spnewtheorem*{theorem*}{Theorem}{\normalshape\bfseries}{\itshape}

\newcommand{\eqdef}{\triangleq}
\renewcommand{\leq}{\leqslant}
\renewcommand{\geq}{\geqslant}
\renewcommand{\epsilon}{\varepsilon}
\newcommand{\rg}[2]{\left \{ #1,\dots,#2  \right \}}

\newcommand{\affect}{\leftarrow}
\newcommand{\raffect}{\xleftarrow{\$}}
\newcommand{\norm}[1]{\left \lvert {#1}  \right \rvert}
\newcommand{\support}[2]{\langle {#2} {\rangle}_{{#1}}}
\newcommand{\sphere}{\mathbb{S}}

\newcommand{\F}{\mathbb{F}}
\newcommand{\N}{\mathbb{N}}
\renewcommand{\P}{\mathbb{P}}
\newcommand{\R}{\mathbb{R}}

\newcommand{\fq}{\F_q}
\newcommand{\fqm}{\F_{q^m}}

\newcommand{\set}[1]{\mathcal{#1}}
\newcommand{\card}[1]{\left \lvert  {#1} \right \rvert }
\newcommand{\bigcard}[1]{\Big \lvert  {#1} \Big \rvert }
\newcommand{\GL}{\operatorname{GL}}
\newcommand{\MS}[3]{#3^{#1 \times #2}}
\newcommand{\grassman}[3]{\boldsymbol{\mathrm{Gr}}_{#1}(#2,#3)}
\DeclareMathOperator{\prob}{\P}
\DeclareMathOperator{\SD}{\Delta}

\newcommand{\word}[1]{\boldsymbol{\mathrm{#1}}}
\newcommand{\av}{\word{a}}
\newcommand{\bv}{\word{b}}
\newcommand{\ev}{\word{e}}
\newcommand{\hv}{\word{h}}
\newcommand{\kv}{\word{k}}
\newcommand{\sv}{\word{s}}
\newcommand{\uv}{\word{u}}
\newcommand{\vv}{\word{v}}
\newcommand{\wv}{\word{w}}
\newcommand{\xv}{\word{x}}
\newcommand{\yv}{\word{y}}

\newcommand{\mat}[1]{\boldsymbol{\mathrm{#1}}}
\newcommand{\Am}{\mat{A}}
\newcommand{\Bm}{\mat{B}}
\newcommand{\Cm}{\mat{C}}
\newcommand{\Dm}{\mat{D}}
\newcommand{\Em}{\mat{E}}
\newcommand{\Gm}{\mat{G}}
\newcommand{\Hm}{\mat{H}}
\newcommand{\Mm}{\mat{M}}
\newcommand{\Rm}{\mat{R}}
\newcommand{\Sm}{\mat{S}}
\newcommand{\Um}{\mat{U}}
\newcommand{\Vm}{\mat{V}}
\newcommand{\Wm}{\mat{W}}
\newcommand{\Xm}{\mat{X}}

\newcommand{\onev}{\mathds{1}}

\DeclareMathOperator{\dec}{\Phi}
\newcommand{\algo}[1]{\mathcal{#1}}
\newcommand{\pk}{\mathsf{pk}}
\newcommand{\tk}{\mathsf{tk}}
\newcommand{\gen}{\mathsf{gen}}
\newcommand{\eval}{\mathsf{eval}}
\newcommand{\invert}{\mathsf{invert}}

\newcommand{\RD}{\mathsf{RD}}
\newcommand{\RSL}{\mathsf{RSL}}

\begin{document}                                

\title{Injective Rank Metric Trapdoor Functions with Homogeneous Errors}	
\author{ \'Etienne Burle\inst{1} \and Philippe Gaborit\inst{2} \and Younes Hatri\inst{1} \and Ayoub Otmani\inst{1}}
\institute{LITIS, University of Rouen Normandie, Normandie Univ, France
\and
XLIM, Université de Limoges, France
\email{\{Etienne.Burle,Ayoub.Otmani,Younes.Hatri\}@univ-rouen.fr}\\
\email{gaborit@unilim.fr}
}
	
\maketitle	

\begin{abstract}

In rank-metric cryptography,  a  vector  from a finite dimensional linear space over a finite field is viewed as the linear space spanned by its entries.
The rank decoding problem which is the  analogue  of the  problem of decoding a random linear code  consists  in recovering a basis of a random noise vector that was used to perturb a set of random linear equations sharing a secret solution. Assuming  the intractability of  this problem, we introduce  a new construction of injective one-way trapdoor functions. Our  solution  departs from the frequent way of building public key primitives from error-correcting codes where, to establish the security, ad hoc assumptions about a hidden structure are made.  
Our method produces a hard-to-distinguish linear code together with low weight vectors which constitute the secret that helps recover the inputs. 
The key idea is to focus on trapdoor functions that take sufficiently enough input vectors sharing the same support. Applying then the error correcting algorithm designed for Low Rank Parity Check (LRPC) codes, we obtain an inverting algorithm that recovers the inputs with overwhelming probability. 
	
\end{abstract}

\section{Introduction}

Trapdoor  functions are informally  cryptographic primitives that compute efficiently images but do not allow the recovering of pre-images unless a secret quantity called the trapdoor is known.  
This basic security property is also called one-wayness.
A classical application of trapdoor functions is the construction of public-key encryption scheme \cite{BeRo93}. However these two notions are not equivalent because not every public-key encryption scheme derives from a trapdoor function. 
This distinction comes from the fact that trapdoor functions require to recover the entire input while encryption functions only have to recover the plaintext, leaving out the randomness if need be. 
Trapdoor functions form an important building tool for achieving the standard notion of CCA2 security when we are concerned with confidentiality under active attacks. 
A generic transformation was proposed in \cite{HKW20} to obtain CCA2 security from an injective trapdoor function. 
There exist currently several trapdoor functions from different algorithmic assumptions: factoring hardness \cite{RSA78}, difficulty of decoding a random linear code \cite{M78}, DDH and LWE assumptions \cite{PW08} and Computational Diffie-Hellman assumption \cite{GH18}.  

In this work, we  are interested in  building  rank-based trapdoor functions.
Rank-based cryptography is close in spirit to code-based cryptography with the difference that the ambient linear space defined over a finite field is endowed with  the rank metric, namely vectors are viewed as matrices. The support of vector is then the linear space spanned by its entries.
The fundamental hard-to-solve problem is the Rank decoding problem which is directly inspired by the decoding problem of random linear codes with the Hamming distance. This latter problem is the core assumption on which relies the security of the McEliece cryptosystem \cite{M78}.
Loosely speaking, the rank decoding problem  consists  in recovering a basis of a random ``noise'' vector that was used to perturb a set of random linear equations sharing a secret solution.

The first rank-metric public key encryption scheme was proposed in \cite{GPT91} by Gabidulin, Paramonov and Tretjakov.
This scheme  which can be seen as an analogue of the McEliece cryptosystem \cite{M78} is based on the class of Gabidulin codes \cite{G85}. 
Gibson showed the weakness of the system through a series of successful attacks \cite{G95,G96} exploiting the algebraic structure 
of Gabidulin codes.
Following these failures, new reparations were published \cite{GO01,GOHA03}, later broken \cite{O05,O05a,O08} and then again  other variants appeared\footnote{See  \cite{FL05,G08,GRH09,RGH10,RGH11,AGHRZ19a,L17,WPR18,LLP20}.}. 
But still new attacks are devised \cite{OTN18,GOT18,BC21}.

This break-and-then-repair cycle follows from the fact that the designers try to hide the secret structure with techniques that are not based on standard assumptions. Currently, the rank decoding problem and its generalization, the Rank Support Learning (RSL) problem \cite{GHPT17a}, are believed to be hard and serve as building blocks for proving the security of the encryption schemes  in \cite{ABDGHRTZABBBO19,AABBBBCDGHZ20,W19}.
However the schemes in \cite{ABDGHRTZABBBO19,AABBBBCDGHZ20} require to use ideal codes and therefore relies on the stronger assumption that the rank decoding problem restricted to structured codes is intractable. In addition, the scheme in \cite{ABDGHRTZABBBO19} has to hide the algebraic structure of Low Rank Parity-Check (LRPC) codes \cite{AGHRZ19a}.

\subsubsection*{Our contribution.}

We introduce  a new construction of a  family of rank metric trapdoor functions provably  hard to invert assuming  the intractability of  the RSL problem. As in \cite{W19},
our solution  departs from the recurrent way of building public key primitives from error-correcting codes that makes non-standard assumptions about a hidden   structure  \cite{M78,CFS01,AABBBBDGGGMMPSTZ21,ABDGHRTZABBBO19}, which in some cases turn out to be false \cite{FGOPT11,FGOPT13,MT23} even leading to cryptanalysis \cite{GOT12,GOT12a,CGGOT13,CGGOT14,OT15,COT14,COT17,BMT23,CMT23}.  
The technique we propose is reminiscent of Alekhnovich's \cite{A03} and Ajtai's \cite{Ajtai96} constructions. \cite{Ajtai96}  shows how to sample an essentially uniform integer matrix $\Am$ along with short trapdoor set of vectors that are orthogonal to $\Am$. Similarly, our mechanism generates a (public) matrix $\Gm \in \MS{k}{(n+L)}{\fqm}$ that is statistically indistinguishable from a uniformly sampled matrix along with a trapdoor matrix $\Wm \in \MS{n}{(n+L)}{\fqm}$ where the rows of $\Wm$ are of  relatively small weight and such that $\Gm \Wm^\mathsf{T} = \mat{0}$. Moreover, under the assumption of the hardness of the (decision) Rank decoding problem, we also manage to get $\Gm$ computationally indistinguishable from a random matrix in the way of \cite{A03} with the Hamming metric.
For a security parameter $\lambda \in \N$, we then introduce a family of injective functions $\set{F}_{\lambda} \subseteq \left\{ f_{\Gm} \mid \Gm \in   \MS{k}{(n+L)}{\fqm} \right\}$ such that $f_{\Gm}(\Xm,\Em) \eqdef \Xm \Gm + \Em$ where $\Xm$ is an arbitrary matrix from $\MS{N}{k}{\fqm}$ and $\Em \in \MS{N}{(n+L)}{\fqm}$ is  an \emph{homogeneous matrix of weight $t$}, that is to say all the entries of $\Em$ generate over $\fq$ a linear subspace of $\fqm$ of dimension $t$. 
As in \cite{W19,AADGZ22,ADGRW22} which use the ``multiple syndromes approach'', the core idea to invert $f_{\Gm}$ is  to take input vectors $(\Xm,\Em)$ with sufficiently enough rows. Adapting then the error-correcting algorithm designed specifically for Low Rank Parity Check (LRPC) codes \cite{AGHRZ19a,ABDGHRTZABBBO19}, it is possible  with overwhelming probability to efficiently recover the inputs. 
We recall that LRPC codes are defined through  homogeneous parity-check matrices and were introduced to solve the rank decoding problem. In our context, the task of inverting $f_{\Gm} $ boils down to solving an instance of the RSL problem. For that purpose  we introduce the concept of semi-homogeneous parity-check matrices which is a generalization of LRPC codes. Thanks to the analysis described in \cite{BO23}, we also give in Theorem~\ref{th:decoding} an upper-bound on the probability that our algorithm fails.

\section{Preliminary Definitions} \label{sec:def}

\subsection{Notation} The  symbol  $\eqdef$ is used to define  the left-hand side object. $\card{\set{S}}$ defines  the cardinality of a set $\set{S}$.
We write $x \raffect \set{S}$ to express that $x $ is sampled  according to the uniform distribution over  a set  $\set{S}$.  
We use the notation $\prob \{ E(x)  \; \vert \; x \raffect \set{S}  \}$ to give the probability that an event $E(x)$ occurs under the constraint that $x \raffect \set{S} $.

The finite field with $q$ elements where $q$ is a power of a prime number is written as $\fq$.
All vectors are regarded by default as row vectors and  denoted by boldface letters like $\av = (a_1,\dots{},a_n)$. 
The linear space over a field $\F$ spanned by vectors $\bv_1,\dots,\bv_k$ belonging to a vector space over a field containing $\F$ is written as $\support{\F}{\bv_1,\dots,\bv_k}$.
For $f \in \F$ and $\set{U} \subseteq \F$, the set $\{ f u \mid u \in \set{U} \}$ is denoted by $f \cdot \set{U}$.
Given two arbitrary sets $\set{A}$, $\set{B}$ included in $\fqm$ where $m \geq 1$, we let $\set{A} \cdot \set{B} \eqdef \support{\fq}{a b \mid a \in \set{A}, b \in \set{B}}$.
The set of $r \times n$ matrices with entries in a set $\set{V} \subseteq \F$ is denoted by $\MS{r}{n}{\set{V}}$ and $\GL_n(\set{V})$ is the subset of $\MS{n}{n}{\set{V}}$  of invertible matrices.
The transpose operation  is denoted by the symbol $^{\mathsf{T}}$. 
For matrices $\Am$ and $\Bm$ having the same number of rows, $\begin{bmatrix}\Am & \vert & \Bm \end{bmatrix}$ represents the matrix obtained by concatenating  the columns of $\Am$ followed by the columns of $\Bm$. 
  
\subsection{Rank Metric}

We consider a finite field extension $\fqm/\fq$ of degree $m \geq 1$.
The \emph{support} of $\xv$ from $\fqm^L$ denoted by $\support{\fq}{\xv}$ is the vector subspace over $\fq$ spanned by its entries namely $\support{\fq}{\xv} \eqdef \support{\fq}{x_1,\dots{},x_L} \subseteq \fqm$. 
The \emph{rank weight} of $\xv$ is  $\norm {\xv} \eqdef \dim \support{\fq}{\xv}$.
Likewise, the support $\support{\fq}{\Xm}$ of a matrix $\Xm = \begin{bmatrix} x_{i,j} \end{bmatrix}$ is the vector subspace over $\fq$ spanned by all its entries, 
and its weight is $\norm {\Xm} \eqdef \dim \support{\fq}{\Xm}$. 
We let $\grassman{w}{q}{m}$ be the set of all $w$-dimensional linear subspaces over $\fq$ included in $\fqm$, and the \emph{sphere} in $\fqm^L$ of radius $w$ centered at $\word{0}$ is denoted by $\sphere_{w}\left(\fqm^L\right)$. We recall that the cardinality of $\grassman{w}{q}{m}$  is  the Gaussian coefficient,
\begin{equation*}
	\bigcard{\grassman{w}{q}{m}} = \prod_{i=0}^{w-1} \frac{q^{m} - q^i}{q^{w} - q^i}.
\end{equation*}
The cardinality of $\sphere_{w}\left(\fqm^L\right ) $  is equal to the number of $m \times L$ $q$-ary matrices of rank  $w$ which is \cite{L06}
\begin{equation}\label{card:sphere}
	\card{\sphere_{w}\left(\fqm^L\right )} = \bigcard{\grassman{w}{q}{m}} \; \prod_{i=0}^{w-1} \left(  q^{L} - q^i \right).
\end{equation}
We can bound (see Lemma \ref{lem:upperboundsphere} in  Appendix \ref{appendix:aux}) the cardinality of a sphere as follows provided that $w + 3 \leq \min\{L , m \}$,  
\begin{equation}\label{eq:boundsphere}
q^{(L+m)w - w^2} 
\;\leq \;
 \card{\sphere_{w}\big(\fqm^L\big)}  
 \;\leq \; 
 e^{2/(q-1)} \; q^{(L+m)w - w^2}
\end{equation}

\begin{definition}[Semi-homogeneous matrix]
	A  matrix $\Mm$ with $\ell$ rows  is \emph{semi-homogeneous with supports} $\set{W}_1,\dots{},\set{W}_\ell$ if the support of its $i$th row is $\set{W}_i \in \grassman{w_i}{q}{m}$ where $w_1,\dots,w_\ell$ are positive integers. A semi-homogeneous matrix is of weight $w$ if all $w_i$ equal $w$.
	When the supports $\set{W}_i$ are all equal to the same linear space $\set{W} \in \grassman{w}{q}{m}$ then $\Mm$ is  \emph{homogeneous} of weight $w$ and support~$\set{W}$. 
\end{definition}

\subsection{Statistical  Indistinguishability and Pseudo-Randomness}

For  random variables $X$ and $Y$ taking values in a finite domain $\set{D}$, the \emph{statistical distance} $\SD\big(X : Y\big)$ between $X$ and $Y$ is
\[
\SD\big(X : Y\big) \eqdef \frac{1}{2} \sum_{w \in \set{D}}\Big \lvert \prob\big\{ X = w \big\}  - \prob\big\{Y = w \big\} \Big \rvert.
\]
A random variable $X$  taking values on  $\set{A}$  is said to be \emph{$\epsilon$-close to   uniform over $\set{A}$} (or  \emph{$\epsilon$-uniform over $\set{A}$})  where $\epsilon$ is a positive real number if its statistical distance from the uniform distribution is at most $\epsilon$. 
Two collections of random variables $\left ( X_n\right)_{ n \in \N }$ and $\left ( Y_n \right) _{  n \in \N }$ are \emph{statistically close} if the statistical distance $\SD(X_n : Y_n)$ is negligible as $n$ tends to $+\infty$. When $\left ( Y_n \right)_{  n \in \N }$  is a collection of uniform random variables then
$\left( X_n \right)_{ n \in \N } $  is \emph{statistically close to uniform}.
Finally, $\left ( X_n\right)_{ n \in \N }$ and $\left ( Y_n \right) _{  n \in \N }$ of random variables are \emph{computationally indistinguishable} if 
for every probabilistic polynomial time algorithm $\algo{D}$ and all sufficiently large $n$'s, there exists a negligible function $\epsilon$ such that
\[
\Big \lvert \;
\prob \Big \{ \algo{D}(X_n) = 1 \Big \}  
-  
\prob \Big \{ \algo{D}( Y_n) = 1 \Big \} 
\;
\Big  \rvert
< \epsilon(n).
\]
When $\left ( Y_n \right)_{  n \in \N }$ is a collection of uniform random variables then $\left( X_n \right)_{ n \in \N }$ is said to be \emph{pseudo-random}.

\begin{proposition}\label{prop:fdist}
	Let $X$ and $Y$ be two random variables taking values  in $\set{A}$, and $g : \set{A} \rightarrow \set{B}$ be an arbitrary function independent from $X$ and $Y$. Then we have
	\[
	\Delta(g(X),g(Y)) \leq \Delta(X,Y).
	\]
\end{proposition}

\subsection{Universal Hashing}

A  family of hash functions $\Big\{ h_{\kv} : \set{A} \rightarrow \set{B} \; \big \vert\; \kv \in \set{K} \Big\}$ where  $\set{A}$, $\set{B}$, $\set{K} \subset\{0,1\}^*$ are finite sets, is \emph{universal} if for all \emph{distinct} $a$ and  $a'$ from $\set{A}$, the following property holds 
\[
\prob\Big \{ h_{\kv}(a) = h_{\kv}(a') \;\;\big \vert \;\; \kv \raffect \set{K}  \Big \} \leq \frac{1}{\card{\set{B}}}.
\]
A classical and well-known example of universal hash functions is obtained by considering  the mapping $\phi_{\Am} :  \F^n \rightarrow \F^r$  such that for every $\xv \in \F^n$, $ \phi_{\Am}(\xv) \eqdef \xv\Am$ where $\F$ is an arbitrary finite field, and $\Am$ is matrix from  $\MS{n}{r}{\F}$ with $r < n$. 

\begin{proposition}\label{prop:phiuniversal}
	The family of hash functions $\left\{ \phi_{\Am} \; \big \vert\; \Am \in \MS{n}{r}{\F} \right\}$ is universal.
\end{proposition}

\begin{proof}
	Let us consider  $\xv = (x_1,\dots,x_n)$ and  $\yv = (y_1,\dots,y_n)$ in $\F^n$ such that  $\xv\ne\yv$. Without loss of generality, we  assume that $x_1 \neq y_1$. Note that $\phi_{\Am}(\xv) = \phi_{\Am}(\yv) $ is equivalent to $\av_1 = - (x_1 - y_1)^{-1} \sum_{j=2}^n (x_j - y_j) \av_j$ where $ \av_{1},\dots, \av_n$ are the rows of $\Am$.
	We then have
	\begin{align*}
		\prob\left\{ \phi_{\Am}(\xv) = \phi_{\Am}(\yv) \;\; \big \vert \;\; \Am\raffect  \MS{n}{r}{\F} \right \} 
		&= \frac{1}{\card{\F}^r}
	\end{align*}
	which proves the proposition.
	\qed
\end{proof}
With help of  the celebrated Leftover Hash Lemma \cite{HILL99}, universal families of hash functions give a very useful tool to bound the statistical distance between a couple of random variables $(K,h_K(X))$ where $K \raffect \set{K}$, $X \raffect \set{A}$ and a uniformly distributed random variable $(K,U) \raffect \set{K} \times \set{B}$. In this work we use the following theorem which is a generalization of the Leftover Hash Lemma to multiple instances.

\begin{theorem}[Theorem 8.38 in \cite{shoupbook}] \label{th:nxLHL}
	Let $\big\{ h_{\kv} :  \set{A} \rightarrow \set{B}  \; \big \vert\;\kv \in \set{K}   \big\}$ be a universal family of hash functions where $\set{A}$, $\set{B}$,  $\set{K}  \subset\{0,1\}^*$ are finite sets.\\
	The random variable  $\big(K,h_K(X_1),\dots{},h_K(X_n)\big)$ is $\dfrac{n}{2} \sqrt{ \dfrac{ \card{\set{B}} }{ \card{\set{A}} }}$--close to uniform over $\set{K} \times \set{B}^n$ for independent and uniformly distributed random variables $K\raffect \set{K}$, $X_1 \raffect \set{A},\dots,X_n \raffect \set{A}$.
\end{theorem}

\subsection{Injective Trapdoor Function}
 
Consider natural numbers $\ell$, $d$, $m$, $n$ that are polynomial functions of the security parameter $\lambda \in \N$.
An \emph{injective trapdoor} function family  with domain $\{0,1\}^m$ and range $\{0,1\}^n$  is given by three  probabilistic polynomial-time algorithms $\mathcal{T} = (\gen, \eval, \invert)$. The key generation algorithm $(\pk,\tk) \affect \mathcal{T}.\gen(\onev^\lambda)$ takes as input a security parameter $\lambda$ and outputs a public key $\pk \in \{0,1\}^{\ell}$ and a trapdoor key $\tk \in \{0,1\}^{d}$. The evaluation algorithm $\yv \affect \mathcal{T}.\eval(\pk,\xv)$ on input $\pk$ and $\xv \in \{0,1\}^{m}$  outputs  $\yv \in  \{0,1\}^{n}$. The inversion algorithm $\mathcal{T}.\invert(\pk,\tk,\yv)$ takes as input $\yv \in  \{0,1\}^{n}$, $\pk$, the trapdoor key $\tk$, and outputs either an $m$-string $\xv$ or $\bot \notin \{0,1\}^{m}$ to indicate that an error occurred.
Additionally, it is required that the inversion algorithm fulfills the \emph{perfect correctness} meaning that for every pair of keys $(\pk,\tk) \affect \mathcal{T}.\gen(\onev^\lambda)$, we should have
\[
\forall \xv \in \{0,1\}^m,\; \;
\mathcal{T}.\invert(\pk,\tk,\mathcal{T}.\eval(\pk,\xv)) = \xv.
\]
It is possible to relax this constraint  to allow the inversion algorithm to fail  for a negligible fraction of keys. This feature is called the ``almost-all-keys'' injectivity property. 
The trapdoor functions that will be presented in Section \ref{sec:trapdoor} do not satisfy nor this condition neither the perfect correctness because  we will see that for every  $(\pk,\tk)$ there exists a fraction of the inputs that cannot be inverted. We will show that this fraction can be made arbitrary small. As there exist generic methods that transform a trapdoor function with decryption errors  into a secure public-key encryption scheme \cite{BeRo93,DNR04,HHK17}, these issues are deferred to the full version of the paper.
We now finish this part by recalling that a trapdoor function should be computationally hard to invert without the trapdoor key.

\begin{definition}
	An injective trapdoor function family $(\gen, \eval, \invert)$ with domain $\{0,1\}^m$ and range $\{0,1\}^n$ is \emph{one-way} if for every probabilistic polynomial-time adversary $\algo{A}$, there exists a negligible function $\epsilon : \N \rightarrow \R$ such that for all sufficiently large $\lambda \in \N$,
	\[
	\prob\Big \{ \algo{A}\big(\pk,\mathcal{T}.\eval(\pk,\xv) \big) = \xv  \;\; \big \vert \;\; (\pk,\tk) \affect \mathcal{T}.\gen(\onev^\lambda) \; ; \; \xv \raffect    \{0,1\}^m \Big \} <  \epsilon(\lambda).
	\]
\end{definition}
 
\section{Intractability Assumptions} \label{sec:complexity}

We gather in this section classical hard problems from rank-metric cryptography.

\begin{definition}[Rank decoding]  
	Let $q$ be a power of a prime number, and consider natural numbers $m$, $n$, $k < n$, $t < n$.
	The \emph{rank decoding} problem asks to find $\ev$ from the input $(\Rm,\ev\Rm^{\mathsf{T}})$ when $\Rm\raffect\MS{(n-k)}{n}{\fqm}$ and $\ev \raffect \sphere_{t}(\fqm^n)$.
\end{definition}
It is currently widely believed that the rank decoding problem is computationally hard to solve for both classical and quantum computers. Indeed, the generic decoding problem in Hamming metric is well-studied and there exists a reduction in \cite{GZ14_sv} from it to the rank decoding problem. Currently, the nature of the algorithms that solve this problem is either ``combinatorial'' or ``algebraic''. For the rank decoding problem, the most efficient ones are described in \cite{AGHT18} and \cite{BBBGT22} respectively. It is important to note that this problem can be solved if one can recover the support $\support{q}{\ev} \in \grassman{t}{q}{m}$.
As explained in \cite{GRS16}, when an arbitrary basis $\epsilon_1,\dots,\epsilon_t$ of $\support{\fq}{\ev}$ is known, then with high chances one expect to fully compute $\ev$ from $(\Rm,\sv)$ where $\sv \eqdef \ev  \Rm^{\mathsf{T}}$ when $nt < m(n-k)$.
Indeed the coordinates of $\ev = \begin{bmatrix}e_1,\dots,e_n \end{bmatrix}$ can be recovered by writing that $e_j = \sum_{d=1}^t x_{j,d} \epsilon_d$ where each $x_{j,d}$ is viewed as an unknown that lies in $\fq$.
One then picks a basis of $\fqm$ over $\fq$ and uses it to project the linear system $\sv = \ev  \Rm^{\mathsf{T}}$ within it.
It results in a linear system involving $nt$ unknowns, namely the $x_{j,d}$ variables, and $m(n-k)$ linear equations defined with coefficients in $\fq$. 
Consequently when the parameters are appropriately chosen, and assuming the linear equations are random, one expects to get a unique solution to the linear system. This particular feature of the rank metric has led the authors in \cite{GHPT17a} to introduce a new fundamental computational problem called the ``Rank Support 
Learning'' problem.

\begin{definition}[Rank Support Learning ($\RSL$) \cite{GHPT17a}]  
	Let $q$ be a power of a prime number, and consider natural numbers $m$, $n$, $k < n$, $t < n$ and $N$.
	The \emph{Rank Support Learning} problem asks to recover $\set{V}$ from $(\Rm, \Vm \Rm^{\mathsf{T}})$ when $\Rm \raffect \MS{(n-k)}{n}{\fqm}$, $\set{V} \raffect \grassman{t}{q}{m}$ and $\Vm \raffect \MS{N}{n}{\set{V}}$.
\end{definition}
Equivalently, RSL can be defined as the problem where an adversary has access to $\Rm$ and $N$ instances $\vv_1 \Rm^{\mathsf{T}},\dots{}, \vv_N \Rm^{\mathsf{T}}$ and has to recover  $\set{V}  \in \grassman{t}{q}{m}$ such that $\support{q}{\vv_i} \subseteq \set{V}$ for every $i \in \rg{1}{N}$. 
Notice that the matrix $\Vm$ that appears in the definition of the RSL problem is an homogeneous matrix of weight $t$.
It is clear that the RSL problem is not harder than the rank decoding problem. Additionally, the rank decoding problem is a special case of the RSL problem with a single sample $N=1$, hence showing that the RSL problem with $N=1$ and the rank decoding problem are computationally equivalent. But the problem becomes easier when $N$ increases. \cite{GHPT17a} gives a combinatorial polynomial-time algorithm when $N\geq nt$, then a sub-exponential algebraic attack was introduced in \cite{DT18b}  when $N\geq kt$. Recently a new combinatorial polynomial-time algorithm has been devised in \cite{BBBG22} when $N > kt m / (m-t)$. Finally \cite{BB21} presents an algebraic attack, whose complexity is given in \cite{BBBG22}, that can be thwarted when $N  < kt$.
We can conclude from this discussion that the RSL problem is intractable when $N < kt$.

\begin{assumption}\label{RSLassumption}
	Let $m$ be a natural number that is a polynomial function of a  parameter $\lambda \in \N$,  and assume that $q = O(1)$, $n =\Theta(m) $, $k = \Theta(n)$, $t  = \Omega \left (n^{\alpha} \right)$ where $\alpha$ is a real number from $ ]0,1]$ and $N < k t$.
	There exists a negligible function $ \epsilon : \N \rightarrow \R$ such that for every probabilistic polynomial-time algorithm $\algo{A}$, and all sufficiently large $\lambda \in \N$,  
	\[
	\prob\Bigg\{ \algo{A}(  \Rm,  \Vm  \Rm^{\mathsf{T}}) = \set{V} \;\; \Big \vert \;\;\Rm \raffect \MS{(n-k)}{n}{\fqm} \; ; \; \set{V} \raffect \grassman{t}{q}{m}\; ; \;\Vm \raffect \MS{N}{n}{\set{V}}\Bigg\} < \epsilon(\lambda).
	\]
\end{assumption}
This assumption about the hardness of RSL enables us to establish the hardness of the search version of the rank decoding problem.

\begin{proposition}\label{RDassumption}
	Let $m$ be a natural number that depends polynomially from a parameter $\lambda \in \N$, and assume that $q = O(1)$, $n =\Theta(m) $, $k = \Theta(n)$, and $t  = \Omega \left (n^{\alpha} \right)$ where $\alpha$ is a real number from $ ]0,1]$. Let $h_\lambda $ be the mapping from  $\MS{(n-k)}{n}{\fqm} \times  \sphere_{t}\left( \fqm^n \right)$ into $\MS{(n-k)}{n}{\fqm} \times \fqm^{n-k}$ such that for every $\Rm  \in  \MS{(n-k)}{n}{\fqm}$ and $\ev  \in \sphere_{t}\left(\fqm^n\right)$,
	\[
	h_\lambda \left( \Rm,\ev \right) \eqdef  (  \Rm, \ev  \Rm^\mathsf{T} ).
	\]   
	Then if Assumption \ref{RSLassumption} holds, there exists   a negligible function $ \epsilon : \N \rightarrow \R$ such that for every probabilistic polynomial time algorithm $\algo{A}$, and all sufficiently large $\lambda \in \N$,  
	\[
	\prob\Big \{\algo{A}(  \Rm,  \ev  \Rm^\mathsf{T}) \in h_{\lambda}^{-1}\big\{ ( \Rm, \ev  \Rm^\mathsf{T} )\big\} \;\;\Big \vert \;\;\Rm \raffect \MS{(n-k)}{n}{\fqm} \; ; \; \ev \raffect \sphere_{t}(\fqm^n)\Big \} < \epsilon(\lambda).
	\] 
\end{proposition}

We now introduce another seemingly less difficult computational problem which consist in distinguishing in polynomial time between the distributions $(\Dm,\ev\Dm)$ and $(\Dm,\uv)$ when $\Dm \raffect \MS{n}{(n-k)}{\fqm}$, $\ev \raffect \sphere_{t}(\fqm^n)$ and $\uv \raffect \fqm^{n-k}$. 
This problem is the \emph{decision} version of the rank decoding problem, and is currently believed to be intractable. Notice that if the family $\left(h_{\lambda}\right)_{\lambda \in \N}$ is not one-way then $( \Dm,\ev  \Dm)$ and $( \Dm,\uv )$ can be distinguished in polynomial-time. A less obvious result is to prove that the converse also holds, hence establishing the equivalence between both versions of the rank decoding problem. 
This can be achieved by devising a ``search-to-decision'' reduction.
Such a reduction exists in \cite{IN89,FS96} in the context of Hamming-metric binary codes ($q = 2$ and $m =1$). The reduction relies on the  Goldreich-Levin theorem \cite{GL89}. The authors of \cite{GHT16} used this reduction to build rank-metric pseudo-random number generators. 
They gave on that occasion a reduction from the Hamming-metric version of the (decision) decoding problem to its rank-metric counterpart. Besides this result which only holds for binary codes, to the best of our knowledge, there does not exist a search-to-decision reduction valid for any field.  
However it is widely believed that the decision rank decoding problem is intractable as well.
 
\begin{assumption}[Decision Rank Decoding] \label{decision_RD}
	Let $m$ be a natural number that is a polynomial function of a  parameter $\lambda \in \N$, and assume that $q = O(1)$, $n =\Theta(m) $, $k = \Theta(n)$ and $t = \Omega \left (n^{\alpha} \right)$ where $\alpha$ is a real number from $ ]0,1]$.  
	There exists a negligible function $ \epsilon : \N \rightarrow \R$ such that for every probabilistic polynomial-time algorithm $\algo{A}$, and all sufficiently large $\lambda \in \N$, 
	\begin{multline*}
		\Bigg \vert\;\;\prob\Bigg\{ \algo{A}( \Rm, \ev \Rm^\mathsf{T}) = 1 \;\;\; \Big \vert \;\;\;\Rm \raffect \MS{(n-k)}{n}{\fqm} \; ; \; \ev \raffect  \sphere_{t}(\fqm^n) \Bigg\} \\
		-
		\prob\Bigg\{\algo{A}(  \Rm,\uv ) = 1 \;\;\; \Big \vert \;\;\;\Rm \raffect \MS{(n-k)}{n}{\fqm} \; ; \;\uv \raffect \fqm^{n-k}\Bigg\} \;\;\Bigg\vert < \epsilon(\lambda)
	\end{multline*}
\end{assumption}
From this assumption, we can prove that the pseudo-randomness  of  $( \Dm,\ev  \Dm)$  is still true when a polynomially bounded number of samples is considered.

\begin{proposition}[Theorem 3.2.6 in \cite{G01_a}]\label{repeated-pseudo} 
	Under Assumption \ref{decision_RD}, there exists a negligible function $\epsilon : \N \rightarrow \R$ such that for every probabilistic polynomial-time algorithm $\algo{A}$, all sufficiently large $\lambda$'s, and for every natural number $\ell$ that is a polynomial function of $\lambda \in \N$,
	\begin{multline*}
		\Bigg \vert \;\;\prob\Bigg\{\algo{A}\left(  \Rm,  \Em \Rm^\mathsf{T}\right) = 1\; \; \; \Big \vert \; \; \;\Rm \raffect \MS{(n-k)}{n}{\fqm} \; ;\; \Em  \raffect\Big(\sphere_{t}(\fqm^n) \Big)^\ell \Bigg\} \\
		- 
		\prob\Bigg\{\algo{A}(\Rm,\Um ) = 1\; \; \; \Big \vert \; \;\;\Rm \raffect \MS{(n-k)}{n}{\fqm} \; ;\; \Um \raffect \MS{\ell}{(n-k)}{\fqm}\Bigg\}\;\; \Bigg\vert < \epsilon(\lambda)
	\end{multline*}
\end{proposition}

\begin{remark} \label{rem:best_attacks}
	All the previous complexity assumptions can  be equivalently expressed with generator matrices by considering the family of functions  $\left(f_\lambda \right)_{\lambda \in \N}$ where $f_\lambda$ is the mapping from $\MS{k}{n}{\fqm} \times  \fqm^{k} \times \sphere_{t}\left( \fqm^n \right)$ into $\MS{k}{n}{\fq} \times\fqm^{n}$ such that
	$f_\lambda\left({\Gm},\xv,\ev \right) \eqdef \left(\Gm,\xv \Gm + \ev\right)$ for every $\Gm  \in \MS{k}{n}{\fq}$, $\xv \in \fqm^k$ and $\ev \in \sphere_{t}(\fqm^n)$.
	It is then not difficult to see that $\left(f_\lambda \right)_{\lambda \in \N}$ is one-way if and only if $\left(h_\lambda \right)_{\lambda \in \N}$ is one-way. 
\end{remark}

\section{A Decoding Algorithm for Homogeneous Errors}\label{sec:decoding}

This section is devoted to explaining how to efficiently solve the RSL problem and decode with a semi-homogeneous parity-check matrix. 
Throughout this section we consider a semi-homogeneous matrix $\Hm \in \MS{\ell}{n}{\fqm}$ of weight $w$ and supports $ (\set{W}_1,\dots{},\set{W}_\ell)$ where $\ell < n$, an integer $t$ and a matrix $\Sm \in \MS{\ell}{N}{\fqm}$. The goal is then to find an homogeneous ``error'' matrix $\Em  \in \MS{n}{N}{\fqm}$ such that the following holds,
\[
\begin{cases}
\Sm = \Hm \Em, \\ 
\support{\fq}{\Em} \in \grassman{t}{q}{m}.
\end{cases}
\]
We provide a two-step procedure denoted by $\dec$ that solves this problem with overwhelming probability provided that $n \leq \ell w$ and $tw\leq N$.
Throughout this section we assume that a  basis $f^{(r)}_1,\dots,f^{(r)}_w$ of $\set{W}_r$ was picked for each $r \in \rg{1}{\ell}$. We will also denote by $\sv_r$ the $r$-th row of $\Sm$.
The main goal of the first step (Algorithm \ref{STEP1}) is to compute a basis of $\support{\fq}{\Em}$. 
The idea is to find an index $r$ in $\rg{1}{\ell}$ such that $\set{E}_r \eqdef \bigcap\limits_{i = 1}^w \left(f^{(r)}_i\right)^{-1} \cdot \support{\fq}{\sv_r}$ belongs to $\grassman{t}{q}{m}$.
If such $r$ exists then we have the equality $\set{E}_r = \support{\fq}{\Em}$. This step then terminates by computing a basis $\epsilon_1,\dots, \epsilon_t$ of $\support{\fq}{\Em} = \set{E}_r$. 

\begin{algorithm}
	\begin{algorithmic}[1]
		\STATE{$r \affect1$}
		\STATE{$\set{E}  \affect \{ \word{0} \}$}  
		\WHILE{$\big(\dim \set{E}  \neq t\big)$  \AND $\big(r \leq \ell\big)$}
		\STATE{$\set{E}  \affect \bigcap\limits_{i = 1}^w  \left(f^{(r)}_i\right)^{-1}  \cdot \support{\fq}{\sv_r}$}
		\hfill \COMMENT{$\sv_r$ is the $r$th row of $\Sm$}
		\STATE{$r \affect r + 1$}
		\ENDWHILE
		\STATE{$\set{B} \affect \emptyset$}	
		\IF{$\dim \set{E} = t$}
		\STATE{$\set{B} \affect \{ \epsilon_1,\dots, \epsilon_t\}$ where $\epsilon_1,\dots, \epsilon_t$ is a basis of $\set{E}$}
		\ENDIF  
		\RETURN{$\set{B}$}
	\end{algorithmic}
	\caption{\textsf{Step I of $\dec(\Hm,\Sm)$ -- Recovering the support.}}\label{STEP1}
\end{algorithm}

Next the second step aims at fully recovering the entries of the matrix $\Em$. It starts by checking whether for each $r$ in $\rg{1}{\ell}$ the dimension of $\set{E} \cdot \set{W}_r$ over $\fq$ is equal to $tw$. When this happens then as a basis of $\set{E} \cdot \set{W}_r$ is given by $f^{(r)}_i \epsilon_j$ with $i\in \rg{1}{w}$ and $j \in \rg{1}{t}$, each entry of $\Sm = \begin{bmatrix} s_{r,c} \end{bmatrix}$ is written as $s_{r,c} = \sum_{i,j} \sigma^{(r,c)}_{i,j} f^{(r)}_i \epsilon_j$ where $\sigma^{(r,c)}_{i,j}$ lies in $\fq$.
Similarly each entry of $\Hm = \begin{bmatrix} h_{r,d} \end{bmatrix}$ with $d \in \rg{1}{n}$ is decomposed as $h_{r,d} = \sum_{i} \nu^{(r,d)}_{i} f^{(r)}_i$ with $\nu^{(r,d)}_{i} $ in $\fq$. Lastly each entry $e_{d,c}$ of the unknown matrix $\Em$ is written as $e_{d,c} = \sum_{j} x^{(d,c)}_{j}\epsilon_j$ where $x^{(d,c)}_{j}$  are unknowns that are sought in $\fq$ so that we have
\begin{align*}
	s_{r,c} =\sum_{d=1}^n h_{r,d} e_{d,c} 
	&= \sum_{d=1}^n \left( \sum_{i=1}^w \nu^{(r,d)}_{i} f^{(r)}_i \right) \left( \sum_{j=1}^t x^{(d,c)}_{j}   \epsilon_j \right)\\
	&= \sum_{i=1}^w \sum_{j=1}^t \left(  \sum_{d=1}^n \nu^{(r,d)}_{i} x^{(d,c)}_{j} \right)  f^{(r)}_i \epsilon_j.
\end{align*}
The latter equality implies that for every $c \in \rg{1}{N}$, we have a system of $\ell tw$ linear equations involving $tn$ unknowns composed of the linear relations 
\[ 
\sigma^{(r,c)}_{i,j} = \sum_{d=1}^n \nu^{(r,d)}_{i} x^{(d,c)}_{j}
\]
where $(r,i,j)$ runs through $\rg{1}{\ell} \times \rg{1}{w} \times\rg{1}{t}$.
As we have taken $\ell w \geq n$ and since $\dim \set{E} \cdot \set{W}_r = tw$  we are sure  to  get a unique solution.   

However, the decoding algorithm $\dec$ we described previously may fail for different reasons. During the first step, $\dec$ will not find $\support{\fq}{\Em}$ if one of the following two events occur:
\begin{enumerate}
	\item  either $\support{\fq}{\sv_r}$ is not equal to $\set{E} \cdot \set{W}_r$  for every $r \in \rg{1}{\ell}$ which always happens when $N<wt$	
	\item  or, for every $r$ such that the equality $\support{\fq}{\sv_r} = \set{E} \cdot \set{W}_r$ holds, we nevertheless have $\set{E} \; \subsetneq \; \bigcap_{i=1}^w \left(f^{(r)}_i\right)^{-1} \cdot \support{\fq}{\sv_r}$.
\end{enumerate}
Lastly, the decoding algorithm may fail during the second step, that is $\Em$ is not recovered, if there exists at least one $r$ in $ \rg{1}{\ell}$ such that the dimension of $\set{E} \cdot \set{W}_r$ is not equal to $tw$. 
We provide in Theorem~\ref{th:decoding} an upper-bound of the decoding failure probability of  $\dec$.

\begin{theorem}\label{th:decoding}
	Consider natural numbers $w$, $t$, $m$, $n$, $\ell$, $N$ such that $\ell < n \leq \ell w$, $(2w - 1) t < m$  and $N \geq tw$.
	For a randomly drawn  semi-homogeneous parity-check matrix $\Hm \in \MS{\ell}{n}{\fqm}$ with supports $\set{W}_1 \raffect \grassman{w}{q}{m},\dots{},\set{W}_\ell \raffect \grassman{w}{q}{m}$ and a random error matrix $\Em \raffect \MS{n}{N}{\set{E}}$ with $\set{E} \raffect \grassman{t}{q}{m}$, the probability that $\dec(\Hm,\Hm\Em) \neq \Em$ is at most $ \prob_{\rm I} +  \prob_{\rm II}$ where
	\[
	\begin{cases}
	\prob_{\rm I} &\; \leq \; \displaystyle \left( 1 - \prod_{i=0}^{tw-1} \left(1 - q^{i-N} \right) + \dfrac{q^{(2w-1)t}}{q^m - q^{t-1}} \right)^\ell, \\
	\prob_{\rm II} &\;\leq \; 1 - \left( 1 - \dfrac{q^{tw}}{q^m - q^{t-1}} \right)^\ell.
	\end{cases}
	\]
\end{theorem}

\begin{proof}
	See Appendix \ref{appendix:failure_decoding}.
\end{proof}

\section{Injective Trapdoor Functions} \label{sec:trapdoor}

We describe in this section our constructions of 
injective trapdoor functions $\set{F}_{\lambda} \subseteq \left\{ f_{\Gm} \mid \Gm \in   \MS{k}{(n+L)}{\fqm} \right\}$
that benefit from the  hardness of the  RSL problem as recalled in Assumption \ref{RSLassumption}. Throughout this section we assume that $q = O(1)$ is a power of prime number and $k$, $n$, $L$, $m$, $t$, $w$ are natural numbers that are supposed to be polynomial functions of a security parameter $\lambda \in \N$ so that Assumption~\ref{decision_RD} holds. We also take  $N \geq t w $ and $nw\geq n+L$. 

\begin{algorithm}
	\begin{algorithmic}[1]
		\FOR{$i \in \rg{1}{n}$} 
			\STATE {$\set{W}_i \raffect \grassman{w}{\fqm}$} 
		\ENDFOR     
    	\STATE{$\Wm_1\raffect \MS{n}{L}{\fqm}$ where $\Wm_1$ is a semi homogeneous with supports $(\set{W}_1,\dots{},\set{W}_n)$}
     	\STATE{$\Wm_2 \raffect \GL_n(\fqm)$  where $\Wm_2$ is a semi homogeneous with supports $(\set{W}_1,\dots{},\set{W}_n)$}
     	\STATE{$\tk \eqdef \begin{bmatrix} \Wm_1 & \vert & \Wm_2 \end{bmatrix}$}
     	\STATE{$\Rm \raffect \fqm^{k \times L}$}
		\STATE{$\pk \eqdef \begin{bmatrix} \Rm & \vert & - \Rm \Wm_1^\mathsf{T} (\Wm_2^{-1})^\mathsf{T} \end{bmatrix}$}
		\RETURN{$(\tk,\pk)$}
	\end{algorithmic}
	\caption{$(\tk,\pk) \affect \gen(\onev^\lambda)$} \label{KG}
\end{algorithm}

The algorithm $\gen$ of $\set{F}_{\lambda}$ is described in Algorithm~\ref{KG}. 
The interesting property of $\gen$ is to output a public key $\pk$ that is a generator matrix $\Gm \in \MS{k}{(n+L)}{\fqm}$ which is (computationally or statistically) indistinguishable from a randomly drawn matrix  from $\MS{k}{(n+L)}{\fqm}$, together with a trapdoor key $\tk$ that is a semi-homogeneous matrix $\Wm \eqdef \begin{bmatrix} \Wm_1 & \vert & \Wm_2 \end{bmatrix}$ in $\MS{n}{(n+L)}{\fqm}$ of weight $w$ such that $\Gm \Wm^{\mathsf{T}} =\mat{0}$. Notice that the definition of $\eval$ is straightforward as it computes $\eval (\pk, \Xm, \Em) \eqdef \Xm \Gm + \Em$ where $\Xm \in \MS{N}{k}{\fqm}$ and $\Em \in \MS{N}{(n+L)}{\fqm}$ is homogeneous of weight $t$. The algorithm $\invert $ takes as input $\Cm \eqdef \Xm \Gm + \Em$, and then applies the decoding algorithm $\dec(\Wm,\Wm\Cm^\mathsf{T})$ that is described in Section~\ref{sec:decoding}. Note that $\Wm$ is not a parity-check matrix of the code generated by the public generator matrix $\Gm$. 
But as $\Wm$ is semi-homogeneous matrix of weight $w$ such that $\Gm \Wm^{\mathsf{T}} =\mat{0}$ and we have taken $N \geq t w$ and $nw\geq n+L$, we can still recover $(\Xm,\Em)$ with overwhelming probability. 
All these previous facts are gathered in the following.

\begin{theorem}
	Let $\set{W}_1 \raffect \grassman{w}{q}{m},\dots{},\set{W}_\ell \raffect \grassman{w}{q}{m}$, and $\Rm$ be uniformly random matrix from $\MS{k}{L}{\fqm}$. For a randomly chosen semi-homogeneous matrix
	$\Wm = 
	\begin{bmatrix} \Wm_1 & \mid & \Wm_2  \end{bmatrix}$ from $\MS{n}{(n+L)}{\fqm}$ with supports $\set{W}_1,\dots{},\set{W}_n$ where $\Wm_1\raffect \MS{n}{L}{\fqm}$ and $\Wm_2 \raffect \GL_n(\fqm)$, the random matrix from $\MS{k}{(n+L)}{\fqm}$ defined as
	$\Gm \eqdef \begin{bmatrix} \Rm & \vert & -  \Rm \Wm_1^\mathsf{T} \left(\Wm_2^\mathsf{T}\right)^{-1} \end{bmatrix}$ satisfies the following properties,
	\begin{enumerate}
		\item $\Gm \Wm^\mathsf{T} = \mat{0}$.
	
		\item Under Assumption~\ref{decision_RD}, $\Gm$ is computationally indistinguishable from a uniformly random  matrix.
		
		\item $\Gm$ is $\epsilon$-close to uniform over $\MS{k}{(n+L)}{\fqm}$ where $\epsilon \eqdef \frac{n}{2} \sqrt{q^{m k - (m+L)w + w^2}}$.
	\end{enumerate}	
\end{theorem}

\begin{proof}
	The fact that $\Gm \Wm^\mathsf{T} = \mat{0}$ is straightforward. Furthermore, $\Gm$ is computationally indistinguishable from a uniformly random matrix if and only if $\Gm_0 \eqdef \begin{bmatrix} \Rm & \vert & - \Rm \Wm_1^\mathsf{T} \end{bmatrix}$ is so. But $\Wm_1$ is a semi-homogeneous matrix of weight $w$, and consequently we can use Proposition~\ref{repeated-pseudo} and claim that $\Gm_0$ is pseudo-random.

	Let us now consider a uniform random matrix $\Um \raffect \MS{k}{(n+L)}{\fqm}$. By Proposition~\ref{prop:fdist} we have

	\[\SD \left( 
	\begin{bmatrix} \Rm \vert -  \Rm \Wm_1^\mathsf{T} \left(\Wm_2^\mathsf{T}\right)^{-1} \end{bmatrix}, \Um\right) 
	\leq 
	\SD \left(\begin{bmatrix} \Rm  \vert  -  \Rm \Wm_1^\mathsf{T}  \end{bmatrix},\Um\right)\]
	Thanks to Proposition \ref{prop:phiuniversal} we can now apply Theorem~\ref{th:nxLHL}. Namely we have the following inequality  
	\begin{align*}
		\SD \left(\begin{bmatrix} \Rm & \vert & - \Rm \Wm_1^\mathsf{T} \end{bmatrix}, \Um\right) & \leq \dfrac{n}{2}\sqrt{\frac{q^{mk}}{\card{\sphere_{w}\left(\fqm^L\right ) }}}.
	\end{align*}
	Using then \eqref{eq:boundsphere} one sees that $\Gm$ is $\epsilon$-close to uniform as claimed.\qed
\end{proof}

\section{Parameters} \label{sec:param} 

We provide in this section the parameters we obtained for different security levels. All our parameters are chosen such that the probability of decoding failure as given in Theorem \ref{th:decoding} is at most $2^{-\lambda}$ where $\lambda$ is the security parameter. Additionally we require that the public matrix $\Gm \eqdef \begin{bmatrix} \Rm & \Big \vert & - \Rm \Wm_1^\mathsf{T} \left(\Wm_2^\mathsf{T}\right)^{-1} \end{bmatrix}$ has to comply with the following security constraints:
\begin{enumerate}	
	\item It is computationally hard to distinguish $\Gm \in \MS{k}{(n+L)}{\fqm}$ from a random matrix. As the best known algorithm for this problem basically solves the search version of the Rank decoding problem, the parameters we compute take only into account the complexity of the best algorithms that solve this problem. We recall that in our case we have to solve $n$ instances of the form $\left(\Rm, \wv^{(1)}_j \Rm^\mathsf{T}\right)$ where $\Rm \in \MS{k}{L}{\fqm}$ and $\wv^{(1)}_j$ is the $j$th row of $\Wm_1 \in \MS{L}{n}{\fqm}$ such that $\support{\fq}{\wv^{(1)}_j} \subseteq \set{W}_j$.
	We denote by $\RD_{q,m}(L,L-k,w)$ the  time complexity of the best algorithm that solves such an instance of the Rank decoding problem. 
	
	\item It is computationally hard to invert $f_{\Gm}(\Xm,\Em) = \Xm \Gm + \Em$ where $\Xm \raffect \MS{N}{k}{\fqm}$ and $\Em \in \MS{N}{(n+L)}{\fqm}$ is a (random) homogeneous matrix of weight $t$. It is clearly an instance of the Rank Support Learning problem with a linear code $\subset \fqm^{n+L}$ of dimension $k$ and $N$ noisy codewords where the support of each error vectors are included in $\support{\fq}{\Em} \in \grassman{t}{q}{m}$. We denote by $\RSL_{q,m}(n+L,k,t,N)$ the time complexity of the best algorithm for solving such an instance. 
	Note that another way to invert $f_{\Gm}$ is to view the problem as solving $N$ instances $\left(\Gm,\xv_i \Gm + \ev_i\right)$ of the Rank decoding problem where $\xv_i$ and $\ev_i$ are respectively the $i$th row of $\Xm$ and $\Em$. The time complexity of the best algorithm for such instances is then $\RD_{q,m}(n+L,k,t)$.  
\end{enumerate}
The goal is then to find parameters such that the following holds
\begin{equation}\label{sec:conditions}
\min \Big \{ \RD_{q,m}(L,L-k,w), \RSL_{q,m}(n+L,k,t,N), \RD_{q,m}(n+L,k,t)  \Big \} \geq 2^\lambda
\end{equation}
Table \ref{tb:comput}  gives examples of parameters for a computationally indistinguishable matrix $\Gm$ at different values of $\lambda$. The last columns also indicates the sizes  in KBytes of the public keys and the ciphertexts.
We took $tw\leq N \leq tw+3$.
We have also computed parameters such that $\Gm$ is $2^{-\lambda}$--close to uniform.
We recall that $\Gm$ is $\epsilon$-close to uniform over $\MS{k}{(n+L)}{\fqm}$ where $\epsilon \eqdef \frac{n}{2} \sqrt{q^{m k - (m+L)w + w^2}}$.
Consequently, besides the constraints that we gave in \eqref{sec:conditions} we now require that $\epsilon < 2^{-\lambda}$.
This last constraint imposes a high value for $w$ that becomes close to $k$ as it can be seen in Table \ref{tb:statis}.

\begin{table}
	\begin{center}	
		\begin{tabular}{*{11}{r}} \toprule
			$\lambda$ (\textsf{Security}) &     $q$ & $m$ & $L$  & $k$ &  $n$ &  $w$ & $t$ & $N$ & \textsf{Public Key (KB)} & \textsf{Ciphertext (KB)} \\ 	\midrule
			$80$  &   $2$  & $179$ & $37$ & $16$ & $163$ & $6$ & $14$ & 84 & 64 & 367  \\ 
			$128$ & $2$ & $293$ & $43$ & $20$ & $261$ & $8$ & $19$ & 153 & 203 & 1,664  \\ 
			$192$ & $2$ & $443$ & $59$ & $27$ & $391$ & $9$ & $26$ & 237 & 618 & 5,694  \\  
			$256$ & $2$ & $409$ & $200$ & $33$ & $521$ & $4$ & $32$ & 128 & 1,134 & 4,608  \\
			\bottomrule
		\end{tabular}
	\end{center}
	\caption{Parameters for $\Gm$ generated by Algorithm \ref{KG} to be computationally indistinguishable}\label{tb:comput}
\end{table}

\begin{table}
	\begin{center}
		\begin{tabular}{*{11}{r}} \toprule
			$\lambda$ (\textsf{Security}) &     $q$ & $m$ & $L$  & $k$ &  $n$ &  $w$ & $t$ & $N$ & \textsf{Public Key (KB)} & \textsf{Ciphertext (KB)} \\ 	\midrule
			$80$  &   $2$  & $499$ & $59$ & $17$ & $163$ & $16$ & $13$ & 208 & 212 & 2,813 \\ 
			$128$ & $2$ & $907$ & $130$ & $21$ & $261$ & $19$ & $20$ & 380 & 860 & 16,450\\
			$192$ & $2$ & $1657$ & $234$ & $29$ & $391$ & $26$ & $28$ & 728 & 3,496 & 92,033\\
			$256$ & $2$ & $2707$ & $129$ & $36$ & $521$ & $35$ & $35$ & 1225 & 7,304 & 263,116\\
			\bottomrule
		\end{tabular}
	\end{center}
	\caption{Parameters for $\Gm$ generated by Algorithm \ref{KG} to be $2^{-\lambda}$--close to uniform}\label{tb:statis}
\end{table}

\section{Conclusion}
The main achievement  of this work is to highlight the first construction of a rank-metric scheme with public keys that are statistically indistinguishable from a random code.
The construction of (injective) trapdoor functions we propose  represents a first step for building a public-key encryption scheme and a Key Encapsulation Mechanism (KEM). The  sizes of the keys are not competitive compared to those in \cite{AABBBBCDGHZ20,ABDGHRTZABBBO19}. 
This difference is due to several reasons. We do not use structured matrices like  ideal codes. Our public keys are computationally hard to distinguish which impact the performances. We impose a probability of decoding failure to be less than $2^{-\lambda}$ where $\lambda$ is the security parameter  and lastly, because the inversion algorithm requires to recover the whole input, our probability of failure is greater than in \cite{ABDGHRTZABBBO19}, especially because of the probability $\prob_{\rm II}\sim \ell q^{tw - m}$ which implies that $m \geq \lambda + t w$. Hence
we do not exclude any drastic key size reduction if we  get rid of the second step that recovers $\Em$ as it is done in \cite{ABDGHRTZABBBO19}. 

\subsubsection*{Acknowledgments.} We are grateful to M. Bardet for providing computer programs that helped us compute complexities of the best existing attacks. E. Burle is supported by RIN100 program funded by R\'egion Normandie. Y. Hatri is supported by RIN Label d'Excellence MINMACS funded by R\'egion Normandie.  P. Gaborit and A. Otmani are supported  by the grant ANR-22-PETQ-0008 PQ-TLS funded by Agence Nationale de la Recherche within France 2030 program. A Otmani is supported by FAVPQC (EIG CONCERT-Japan \& CNRS).

\bibliographystyle{plain}


\begin{thebibliography}{10}

\bibitem{AABBBBDGGGMMPSTZ21}
Carlos {Aguilar Melchor}, Nicolas Aragon, Paulo Barreto, Slim Bettaieb,
  Lo{\"i}c Bidoux, Olivier Blazy, Jean-Christophe Deneuville, Philippe Gaborit,
  Shay Gueron, Tim G{\"u}neysu, Rafael Misoczki, Edoardo Persichetti, Nicolas
  Sendrier, Jean-Pierre Tillich, and Gilles Z{\'e}mor.
\newblock {BIKE}.
\newblock Round 3 Submission to the NIST Post-Quantum Cryptography Call,
  v.~4.2, September 2021.

\bibitem{AABBBBCDGHZ20}
Carlos {Aguilar Melchor}, Nicolas Aragon, Slim Bettaieb, Loïc Bidoux, Olivier
  Blazy, Maxime Bros, Alain Couvreur, Jean-Christophe Deneuville, Philippe
  Gaborit, Gilles Z{\'e}mor, and Adrien Hauteville.
\newblock Rank quasi cyclic {(RQC)}.
\newblock Second Round submission to NIST Post-Quantum Cryptography call, April
  2020.

\bibitem{AADGZ22}
Carlos {Aguilar Melchor}, Nicolas Aragon, Victor Dyseryn, Philippe Gaborit, and
  Gilles Z\'emor.
\newblock {LRPC} codes with multiple syndromes: Near ideal-size kems without
  ideals.
\newblock In Jung~Hee Cheon and Thomas Johansson, editors, {\em Post-Quantum
  Cryptography - 13th International Workshop, PQCrypto 2022, Virtual Event,
  September 28-30, 2022, Proceedings}, volume 13512 of {\em Lecture Notes in
  Computer Science}, pages 45--68. Springer, 2022.

\bibitem{Ajtai96}
Mikl{\'o}s Ajtai.
\newblock Generating hard instances of lattice problems (extended abstract).
\newblock In {\em STOC '96}, 1996.

\bibitem{A03}
Michael Alekhnovich.
\newblock More on average case vs approximation complexity.
\newblock In {\em 44th Symposium on Foundations of Computer Science {(FOCS
  2003)}, 11-14 October 2003, Cambridge, MA, USA, Proceedings}, pages 298--307.
  {IEEE} Computer Society, 2003.

\bibitem{ABDGHRTZABBBO19}
Nicolas Aragon, Olivier Blazy, Jean-Christophe Deneuville, Philippe Gaborit,
  Adrien Hauteville, Olivier Ruatta, Jean-Pierre Tillich, Gilles Z{\'e}mor,
  Carlos {Aguilar Melchor}, Slim Bettaieb, Lo{\"i}c Bidoux, Magali Bardet, and
  Ayoub Otmani.
\newblock {ROLLO} (merger of {Rank-Ouroboros, LAKE and LOCKER}).
\newblock Second round submission to the NIST post-quantum cryptography call,
  March 2019.

\bibitem{ADGRW22}
Nicolas Aragon, Victor Dyseryn, Philippe Gaborit, Pierre Loidreau, Julian
  Renner, and Antonia Wachter-Zeh.
\newblock {LowMS}: a new rank metric code-based kem without ideal structure.
\newblock Cryptology ePrint Archive, Paper 2022/1596, 2022.
\newblock \url{https://eprint.iacr.org/2022/1596}.

\bibitem{AGHRZ19a}
Nicolas Aragon, Philippe Gaborit, Adrien Hauteville, Olivier Ruatta, and Gilles
  Z{\'{e}}mor.
\newblock Low rank parity check codes: New decoding algorithms and applications
  to cryptography.
\newblock {\em IEEE Trans. Inform. Theory}, 65(12):7697--7717, 2019.

\bibitem{AGHT18}
Nicolas Aragon, Philippe Gaborit, Adrien Hauteville, and Jean-Pierre Tillich.
\newblock A new algorithm for solving the rank syndrome decoding problem.
\newblock In {\em 2018 {IEEE} International Symposium on Information Theory,
  {ISIT} 2018, Vail, CO, USA, June 17-22, 2018}, pages 2421--2425. IEEE, 2018.

\bibitem{BB21}
Magali Bardet and Pierre Briaud.
\newblock An algebraic approach to the rank support learning problem.
\newblock In Jung~Hee Cheon and Jean-Pierre Tillich, editors, {\em Post-Quantum
  Cryptography}, volume 12841 of {\em LNCS}, pages 442--462, Cham, 2021.
  Springer International Publishing.

\bibitem{BBBGT22}
Magali Bardet, Pierre Briaud, Maxime Bros, Philippe Gaborit, and Jean-Pierre
  Tillich.
\newblock Revisiting algebraic attacks on minrank and on the rank decoding
  problem, 2022.

\bibitem{BMT23}
Magali Bardet, Rocco Mora, and Jean{-}Pierre Tillich.
\newblock Polynomial time attack on high rate random alternant codes.
\newblock {\em CoRR}, abs/2304.14757, 2023.

\bibitem{BeRo93}
Mihir Bellare and Phillip Rogaway.
\newblock Random oracles are practical: {A} paradigm for designing efficient
  protocols.
\newblock In Dorothy~E. Denning, Raymond Pyle, Ravi Ganesan, Ravi~S. Sandhu,
  and Victoria Ashby, editors, {\em {CCS} '93, Proceedings of the 1st {ACM}
  Conference on Computer and Communications Security, Fairfax, Virginia, USA,
  November 3-5, 1993}, pages 62--73. {ACM}, 1993.

\bibitem{BBBG22}
Lo{\"i}c Bidoux, Pierre Briaud, Maxime Bros, and Philippe Gaborit.
\newblock {RQC} revisited and more cryptanalysis for rank-based cryptography.
\newblock {\em ArXiv}, abs/2207.01410, 2022.

\bibitem{BC21}
Maxime Bombar and Alain Couvreur.
\newblock Decoding supercodes of {G}abidulin codes and applications to
  cryptanalysis.
\newblock In Jung~Hee Cheon and Jean-Pierre Tillich, editors, {\em Post-Quantum
  Cryptography}, pages 3--22, Cham, 2021. Springer International Publishing.

\bibitem{CFS01}
Nicolas Courtois, Matthieu Finiasz, and Nicolas Sendrier.
\newblock How to achieve a {McEliece}-based digital signature scheme.
\newblock In {\em Advances in Cryptology - ASIACRYPT~2001}, volume 2248 of {\em
  LNCS}, pages 157--174, Gold Coast, Australia, 2001. Springer.

\bibitem{CGGOT14}
Alain Couvreur, Philippe Gaborit, Val{\'{e}}rie Gauthier{-}Uma{\~{n}}a, Ayoub
  Otmani, and Jean-Pierre Tillich.
\newblock Distinguisher-based attacks on public-key cryptosystems using
  {Reed-Solomon} codes.
\newblock {\em Des. Codes Cryptogr.}, 73(2):641--666, 2014.

\bibitem{CGGOT13}
Alain Couvreur, Philippe Gaborit, Val{\'e}rie Gautier, Ayoub Otmani, and
  Jean-Pierre Tillich.
\newblock {Distinguisher-Based Attacks on Public-Key Cryptosystems Using
  {R}eed-{S}olomon Codes}.
\newblock In {\em {International Workshop on Coding and Cryptography - WCC
  2013}}, pages 181--193, Bergen, Norway, April 2013.

\bibitem{CMT23}
Alain Couvreur, Rocco Mora, and Jean{-}Pierre Tillich.
\newblock A new approach based on quadratic forms to attack the mceliece
  cryptosystem.
\newblock {\em CoRR}, abs/2306.10294, 2023.

\bibitem{COT14}
Alain Couvreur, Ayoub Otmani, and Jean-Pierre Tillich.
\newblock New identities relating wild {G}oppa codes.
\newblock {\em Finite Fields Appl.}, 29:178--197, 2014.

\bibitem{COT17}
Alain Couvreur, Ayoub Otmani, and Jean-Pierre Tillich.
\newblock Polynomial time attack on wild {M}c{E}liece over quadratic
  extensions.
\newblock {\em IEEE Trans. Inform. Theory}, 63(1):404--427, 1 2017.

\bibitem{DT18b}
Thomas {Debris-Alazard} and Jean-Pierre Tillich.
\newblock Two attacks on rank metric code-based schemes: Ranksign and an
  identity-based-encryption scheme.
\newblock In {\em Advances in Cryptology - ASIACRYPT~2018}, volume 11272 of
  {\em LNCS}, pages 62--92, Brisbane, Australia, December 2018. Springer.

\bibitem{DNR04}
Cynthia Dwork, Moni Naor, and Omer Reingold.
\newblock Immunizing encryption schemes from decryption errors.
\newblock In Christian Cachin and Jan~L. Camenisch, editors, {\em Advances in
  Cryptology - EUROCRYPT 2004}, pages 342--360. Springer Berlin Heidelberg,
  2004.

\bibitem{FGOPT11}
Jean-Charles Faug{\`e}re, Val{\'e}rie Gauthier, Ayoub Otmani, Ludovic Perret,
  and Jean-Pierre Tillich.
\newblock A distinguisher for high rate {McEliece} cryptosystems.
\newblock In {\em Proc. IEEE Inf. Theory Workshop- ITW~2011}, pages 282--286,
  Paraty, Brasil, October 2011.

\bibitem{FGOPT13}
Jean-Charles Faug{\`e}re, Val{\'e}rie Gauthier, Ayoub Otmani, Ludovic Perret,
  and Jean-Pierre Tillich.
\newblock A distinguisher for high rate {McEliece} cryptosystems.
\newblock {\em IEEE Trans. Inform. Theory}, 59(10):6830--6844, October 2013.

\bibitem{FL05}
C{\'{e}}dric Faure and Pierre Loidreau.
\newblock A new public-key cryptosystem based on the problem of reconstructing
  \emph{p}-polynomials.
\newblock In {\em Coding and Cryptography, International Workshop, {WCC} 2005,
  Bergen, Norway, March 14-18, 2005. Revised Selected Papers}, pages 304--315,
  2005.

\bibitem{FS96}
Jean-Bernard Fischer and Jacques Stern.
\newblock An efficient pseudo-random generator provably as secure as syndrome
  decoding.
\newblock In Ueli Maurer, editor, {\em Advances in Cryptology - EUROCRYPT'96},
  volume 1070 of {\em LNCS}, pages 245--255. Springer, 1996.

\bibitem{GRH09}
Ernst Gabidulin, Haitam Rashwan, and Bahram Honary.
\newblock On improving security of {GPT} cryptosystems.
\newblock In {\em Proc. IEEE Int. Symposium Inf. Theory - ISIT}, pages
  1110--1114. IEEE, 2009.

\bibitem{G85}
Ernst~M. Gabidulin.
\newblock Theory of codes with maximum rank distance.
\newblock {\em Problemy Peredachi Informatsii}, 21(1):3--16, 1985.

\bibitem{G08}
Ernst~M. Gabidulin.
\newblock Attacks and counter-attacks on the {GPT} public key cryptosystem.
\newblock {\em Des. Codes Cryptogr.}, 48(2):171--177, 2008.

\bibitem{GO01}
Ernst~M. Gabidulin and Alexei~V. Ourivski.
\newblock Modified {GPT} {PKC} with right scrambler.
\newblock {\em Electron. Notes Discrete Math.}, 6:168--177, 2001.

\bibitem{GOHA03}
Ernst~M. Gabidulin, Alexei~V. Ourivski, Bahram Honary, and Bassem Ammar.
\newblock Reducible rank codes and their applications to cryptography.
\newblock {\em IEEE Trans. Inform. Theory}, 49(12):3289--3293, 2003.

\bibitem{GPT91}
Ernst~M. Gabidulin, A.~V. Paramonov, and O.~V. Tretjakov.
\newblock Ideals over a non-commutative ring and their applications to
  cryptography.
\newblock In {\em Advances in Cryptology - EUROCRYPT'91}, number 547 in LNCS,
  pages 482--489, Brighton, April 1991.

\bibitem{GHPT17a}
Philippe Gaborit, Adrien Hauteville, Duong~Hieu Phan, and Jean{-}Pierre
  Tillich.
\newblock Identity-based encryption from rank metric.
\newblock In {\em Advances in Cryptology - CRYPTO2017}, volume 10403 of {\em
  LNCS}, pages 194--226, Santa Barbara, CA, USA, August 2017. Springer.

\bibitem{GHT16}
Philippe Gaborit, Adrien Hauteville, and Jean{-}Pierre Tillich.
\newblock Ranksynd a {PRNG} based on rank metric.
\newblock In {\em Post-Quantum Cryptography~2016}, pages 18--28, Fukuoka,
  Japan, February 2016.

\bibitem{GOT18}
Philippe Gaborit, Ayoub Otmani, and Herv{\'{e}} Tal{\'{e}}-Kalachi.
\newblock Polynomial-time key recovery attack on the {F}aure-{L}oidreau scheme
  based on {Gabidulin} codes.
\newblock {\em Des. Codes Cryptogr.}, 86(7):1391--1403, 2018.

\bibitem{GRS16}
Philippe Gaborit, Olivier Ruatta, and Julien Schrek.
\newblock On the complexity of the rank syndrome decoding problem.
\newblock {\em IEEE Trans. Inform. Theory}, 62(2):1006--1019, 2016.

\bibitem{GZ14_sv}
Philippe Gaborit and Gilles Z{\'{e}}mor.
\newblock On the hardness of the decoding and the minimum distance problems for
  rank codes.
\newblock {\em IEEE IT}, 2016.

\bibitem{GH18}
Sanjam Garg and Mohammad Hajiabadi.
\newblock Trapdoor functions from the computational diffie-hellman assumption.
\newblock In Hovav Shacham and Alexandra Boldyreva, editors, {\em Advances in
  Cryptology - {CRYPTO} 2018 - 38th Annual International Cryptology Conference,
  Santa Barbara, CA, USA, August 19-23, 2018, Proceedings, Part {II}}, volume
  10992 of {\em Lecture Notes in Computer Science}, pages 362--391. Springer,
  2018.

\bibitem{GOT12}
Val{\'{e}}rie Gauthier, Ayoub Otmani, and Jean-Pierre Tillich.
\newblock A distinguisher-based attack of a homomorphic encryption scheme
  relying on {Reed-Solomon} codes.
\newblock {\em CoRR}, abs/1203.6686, 2012.

\bibitem{GOT12a}
Val{\'{e}}rie Gauthier, Ayoub Otmani, and Jean-Pierre Tillich.
\newblock A distinguisher-based attack on a variant of {McEliece's}
  cryptosystem based on {Reed-Solomon} codes.
\newblock {\em CoRR}, abs/1204.6459, 2012.

\bibitem{G95}
Keith Gibson.
\newblock Severely denting the {Gabidulin} version of the {McEliece} public key
  cryptosystem.
\newblock {\em Des. Codes Cryptogr.}, 6(1):37--45, 1995.

\bibitem{G96}
Keith Gibson.
\newblock The security of the {Gabidulin} public key cryptosystem.
\newblock In Ueli Maurer, editor, {\em Advances in Cryptology - EUROCRYPT '96},
  volume 1070 of {\em LNCS}, pages 212--223. Springer, 1996.

\bibitem{G01_a}
Oded Goldreich.
\newblock {\em The Foundations of Cryptography - Volume 1, Basic Techniques}.
\newblock Cambridge University Press, 2001.

\bibitem{GL89}
Oded Goldreich and Leonid~A Levin.
\newblock A hard-core predicate for all one-way functions.
\newblock In {\em Proceedings of the twenty-first annual ACM symposium on
  Theory of computing}, pages 25--32. ACM, 1989.

\bibitem{HILL99}
Johan H{\aa}stad, Russell Impagliazzo, Leonid~A Levin, and Michael Luby.
\newblock A pseudorandom generator from any one-way function.
\newblock {\em SIAM Journal on Computing}, 28(4):1364--1396, 1999.

\bibitem{HHK17}
Dennis Hofheinz, Kathrin H{\"o}velmanns, and Eike Kiltz.
\newblock A modular analysis of the {F}ujisaki-{O}kamoto transformation.
\newblock In {\em Theory of Cryptography Conference}, pages 341--371. Springer,
  2017.

\bibitem{HKW20}
Susan Hohenberger, Venkata Koppula, and Brent Waters.
\newblock Chosen ciphertext security from injective trapdoor functions.
\newblock In Daniele Micciancio and Thomas Ristenpart, editors, {\em Advances
  in Cryptology -- CRYPTO 2020}, pages 836--866, Cham, 2020. Springer
  International Publishing.

\bibitem{IN89}
Russell Impagliazzo and Moni Naor.
\newblock Efficient cryptographic schemes provably as secure as subset sum.
\newblock In {\em 30th Annual Symposium on Foundations of Computer Science,
  North Carolina, USA, 30 October - 1 November 1989}, pages 236--241. {IEEE}
  Computer Society, 1989.

\bibitem{LLP20}
Julien Lavauzelle, Pierre Loidreau, and Ba-Duc Pham.
\newblock {RAMESSES, a Rank Metric Encryption Scheme with Short Keys}.
\newblock working paper or preprint, January 2020.

\bibitem{L06}
Pierre Loidreau.
\newblock Properties of codes in rank metric, 2006.

\bibitem{L17}
Pierre Loidreau.
\newblock A new rank metric codes based encryption scheme.
\newblock In {\em Post-Quantum Cryptography~2017}, volume 10346 of {\em LNCS},
  pages 3--17. Springer, 2017.

\bibitem{M78}
Robert~J. McEliece.
\newblock {\em A Public-Key System Based on Algebraic Coding Theory}, pages
  114--116.
\newblock Jet Propulsion Lab, 1978.
\newblock DSN Progress Report 44.

\bibitem{MT23}
Rocco Mora and Jean{-}Pierre Tillich.
\newblock On the dimension and structure of the square of the dual of a goppa
  code.
\newblock {\em Des. Codes Cryptogr.}, 91(4):1351--1372, 2023.

\bibitem{OT15}
Ayoub Otmani and Herv{\'{e}} Tal{\'{e}}-Kalachi.
\newblock Square code attack on a modified {S}idelnikov cryptosystem.
\newblock In {\em Codes, Cryptology, and Information Security - First
  International Conference, {C2SI} 2015, Proceedings}, volume 9084 of {\em
  Lecture Notes in Computer Science}, pages 173--183. Springer, 2015.

\bibitem{OTN18}
Ayoub Otmani, Herv{\'{e}} Tal{\'{e}}-Kalachi, and S{\'{e}}lestin Ndjeya.
\newblock Improved cryptanalysis of rank metric schemes based on {G}abidulin
  codes.
\newblock {\em Des. Codes Cryptogr.}, 86(9):1983--1996, 2018.

\bibitem{O05a}
Raphael Overbeck.
\newblock Extending {G}ibson's attacks on the {GPT} cryptosystem.
\newblock In Oyvind Ytrehus, editor, {\em WCC 2005}, volume 3969 of {\em LNCS},
  pages 178--188. Springer, 2005.

\bibitem{O05}
Raphael Overbeck.
\newblock A new structural attack for {GPT} and variants.
\newblock In {\em Mycrypt}, volume 3715 of {\em LNCS}, pages 50--63, 2005.

\bibitem{O08}
Raphael Overbeck.
\newblock Structural attacks for public key cryptosystems based on {Gabidulin}
  codes.
\newblock {\em J. Cryptology}, 21(2):280--301, 2008.

\bibitem{PW08}
Chris Peikert and Brent Waters.
\newblock Lossy trapdoor functions and their applications.
\newblock In Cynthia Dwork, editor, {\em Proceedings of the 40th Annual {ACM}
  Symposium on Theory of Computing, Victoria, British Columbia, Canada, May
  17-20, 2008}, pages 187--196. {ACM}, 2008.

\bibitem{RGH10}
Haitam Rashwan, Ernst Gabidulin, and Bahram Honary.
\newblock A smart approach for {GPT} cryptosystem based on rank codes.
\newblock In {\em Proc. IEEE Int. Symposium Inf. Theory - ISIT}, pages
  2463--2467. IEEE, 2010.

\bibitem{RGH11}
Haitam Rashwan, Ernst Gabidulin, and Bahram Honary.
\newblock Security of the {GPT} cryptosystem and its applications to
  cryptography.
\newblock {\em Security and Communication Networks}, 4(8):937--946, 2011.

\bibitem{RSA78}
Ronald~L. Rivest, Adi Shamir, and Leonard~M. Adleman.
\newblock A method for obtaining digital signatures and public-key
  cryptosystems.
\newblock {\em Commun. ACM}, 21(2):120--126, 1978.

\bibitem{shoupbook}
Victor Shoup.
\newblock {\em A Computational Introduction to Number Theory and Algebra}.
\newblock Cambridge University Press, USA, 2 edition, 2008.

\bibitem{WPR18}
Antonia {Wachter-Zeh}, Sven {Puchinger}, and Julian {Renner}.
\newblock Repairing the {F}aure-{L}oidreau public-key cryptosystem.
\newblock In {\em Proc. IEEE Int. Symposium Inf. Theory - ISIT}, pages
  2426--2430, 2018.

\bibitem{W19}
Li-Ping Wang.
\newblock {Loong}: a new {IND-CCA}-secure code-based {KEM}.
\newblock In {\em 2019 IEEE International Symposium on Information Theory
  (ISIT)}, pages 2584--2588, 2019.

\bibitem{BO23}
Étienne Burle and Ayoub Otmani.
\newblock An upper-bound on the decoding failure probability of the lrpc
  decoder, 2023.
\end{thebibliography}

\appendix

\section{Auxiliary Result} \label{appendix:aux}

\begin{lemma}  \label{lem:upperboundsphere}
	Let us assume that $w + 3 \leq \min\{L , m \}$. Then we have   
	\begin{equation*}
		q^{(L+m)w - w^2} \;\leq \;\card{\sphere_{w}\big(\fqm^L\big)}  \;\leq \; e^{2/(q-1)} \; q^{(L+m)w - w^2}
	\end{equation*}
\end{lemma}

\begin{proof}
	Using the expression of the cardinality of $\sphere_{w}\big(\fqm^L\big)$ from \eqref{card:sphere} we therefore have
	\begin{align*}
		\card{\sphere_{w}\big(\fqm^L\big)} = \prod_{i=0}^{w-1} \left(  q^{L} - q^i \right)  \frac{q^{m} - q^i}{q^{w} - q^i}
		& =  q^{(L+m-w)w} \prod_{i=0}^{w-1} \left( 1 - q^{i-L} \right)   \prod_{i=0}^{w-1}  \frac{1 - q^{i-m}}{1 - q^{i-w}} 
	\end{align*}
	Exploiting the fact that for every $x \in [0,1/2]$, $e^{-2x} \leq 1 -x \leq e^{-x}$ we can write that
	\begin{align*}
		\card{\sphere_{w}\big(\fqm^L\big)} &\geq q^{(L+m-w)w} \prod_{i=0}^{w-1}e^{-2q^{i-L}} \; \prod_{i=0}^{w-1} \frac{e^{-2q^{i-m}}}{e^{-q^{i-w}}}\\
										   &\geq  q^{(L+m-w)w} e^{\left(-2q^{-L} - 2 q^{-m} + q^{-w} \right)\frac{q^w-1}{q-1}}
	\end{align*}
	Let us set $\gamma \eqdef q/(q-1)$. This means that $q^{w-1} \leq \frac{q^w-1}{q-1} \leq \gamma q^{w-1}$ which entails that
	\begin{align*}
		\card{\sphere_{w}\big(\fqm^L\big)} 	
		&\geq q^{(L+m-w)w} \; e^{-2\gamma q^{-L+w -1 } - 2 \gamma q^{-m+w-1} + q^{-1}}
	\end{align*}
	But the conditions $w + 3 \leq L$ and $q \geq 2$ enable us to write that $2\gamma q^{-L+w -1 } \leq 2 \gamma q^{-4} \leq \frac{1}{2q}$. By the same arguments we also have $2\gamma q^{-m+w -1 } \leq  \frac{1}{2q}$. This implies that
	\[
	e^{-2\gamma q^{-L+w -1 } - 2 \gamma q^{-m+w-1} + q^{-1}} \geq 1.
	\]
	We have hence proved that $	\card{\sphere_{w}\big(\fqm^L\big)} \geq q^{(L+m - w  )w}$.
	Next, by similar techniques, we can show that
	\begin{align*}
		\card{\sphere_{w}\big(\fqm^L\big)} 	&\leq  q^{(L+m-w)w} \; e^{\left(-q^{-L} -  q^{-m} + 2 q^{-w} \right)\frac{q^w-1}{q-1}}\\
											&\leq  q^{(L+m-w)w} \; e^{ 2 q^{-w} \frac{q^w-1}{q-1}} \leq  q^{(L+m-w)w} \; e^{ 2/(q-1)}
	\end{align*}
	which concludes the proof of the lemma.
	\qed
\end{proof}

\section{Upper-Bound on the Decoding Failure Probability} \label{appendix:failure_decoding}

We now turn to the question of proving Theorem \ref{th:decoding} by bounding the probability $\prob\Big \{\dec(\Hm, \Hm\Em) \neq \Em \Big\}$ that $\dec$ fails on a random input $(\Hm,\Hm\Em)$. For that purpose we define  by $\prob_{\rm I}$ and $\prob_{\rm II}$ the probability that $\dec$ fails at the first and second step respectively.
We then clearly have $\prob\Big  \{\dec(\Hm, \Hm\Em) \neq \Em\Big\} = \prob_{\rm I} + (1- \prob_{\rm I}) \prob_{\rm II}$ which implies that
$	\prob\Big \{\dec(\Hm, \Hm\Em) \neq \Em \Big\}
	\leq \prob_{\rm I} + \prob_{\rm II}. $
We have seen in Section \ref{sec:decoding} that the decoding algorithm $\dec$ fail during the first step if one of the following two events occur: either $\support{\fq}{\sv_r}$ is not equal to $\set{E} \cdot \set{W}_r$ for every $r \in \rg{1}{\ell}$, or each time we have the equality $\support{\fq}{\sv_r} = \set{E} \cdot \set{W}_r$, the strict inclusion $\set{E} \; \subsetneq \; \bigcap_{i=1}^w  \left(f^{(r)}_i\right)^{-1} \cdot \support{\fq}{\sv_r}$ holds.
As by assumption the rows of $\Hm$ are drawn independently, we see that $\prob_{\rm I}$ is at most 
\begin{equation}\label{eq:P1}
	\prod_{r = 1}^\ell 
	\left( 
	\prob \Big \{   \support{\fq}{\sv_r} \neq \set{E} \cdot \set{W}_r \Big\}  
	+ 
	\prob \left \{ 
	\set{E} \neq \bigcap_{i=1}^w  \left(f^{(r)}_{i}\right)^{-1} \cdot \support{\fq}{\sv_r} \;\; \Big \vert \;\; \support{\fq}{\sv_r} = \set{E} \cdot \set{W}_r\right\} 
	\right). 
\end{equation} 
The last failure case  is although $\support{\fq}{\Em}$ has been correctly computed, it cannot compute the entries of $\Em$ because for at least one $r$ in $\rg{1}{\ell}$, the dimension of $\set{E} \cdot \set{W}_r$ is not equal to $tw$. 
We see that we have 
\begin{equation}\label{eq:PII}
	\prob_{\rm II} = 1 - \prod_{r=1}^\ell \prob\Big\{\dim \; \set{E} \cdot \set{W}_r = tw \Big\}.
\end{equation}
The rest of this section is devoted to proving bounds for $\prob_{\rm I}$ and $\prob_{\rm II}$, mainly using results from LRPC decoding described in \cite{BO23}.

In order  to bound $\prob_{\rm I}$ we bound $\prob \Big \{ \support{\fq}{\sv_r} \neq \set{E} \cdot \set{W}_r \Big\}$ with Proposition \ref{prop:PI} and $\prob \left \{ \set{E} \neq \bigcap_{i=1}^w  \left(f^{(r)}_{i}\right)^{-1} \cdot \support{\fq}{\sv_r}\;\;  \Big \vert \;\; \support{\fq}{\sv_r} = \set{E} \cdot \set{W}_r \right\}$ in Theorem \ref{th:bound:intersection}, and with \eqref{eq:P1} we get the result.
\begin{proposition}[Proposition 3 in \cite{BO23}]\label{prop:PI}
	Assume that $N \geq tw$. For a random homogeneous matrix $\Em \raffect \MS{n}{N}{\set{E}}$ and a random vector $\hv \raffect \set{W}^n$ where $\set{E} \raffect \grassman{t}{q}{m}$ and $\set{W} \raffect \grassman{w}{q}{m}$, the probability that $\support{\fq}{\hv \Em}$ is different from $\set{E} \cdot \set{W}$ is at most
	\[
	\prob \left \{ \support{\fq}{\hv \Em}  \neq \set{E} \cdot \set{W} \right\} \leq 1 - \prod_{i=0}^{tw-1} \left(1 - q^{i-N} \right)  
	\]
\end{proposition}
\begin{theorem}[Theorem 2 in \cite{BO23}]\label{th:bound:intersection}
	Let $\set{U} \eqdef \set{E} \cdot \set{W}$ where $\set{W} \in \grassman{w}{q}{m}$ and $\set{E} \raffect \grassman{t}{q}{m}$ with $(2w-1)t < m$.
	Then  for an arbitrary basis $f_1,\dots{},f_w$ of $\set{W}$, we have
	\begin{equation*}
	\prob 
	\left \{ 
	\set{E} =   \bigcap_{i=1}^w  f_i^{-1}  \cdot \set{U} \;\; \Big \vert \;\;\set{E} \raffect  \grassman{t}{q}{m}
	\right\} 
	\; \geq \;
	1 -   \frac{q^{(2w   - 1)t}}{q^m - q^{t-1}}\cdot
	\end{equation*}
\end{theorem}
For $\prob_{\rm II}$, we bound $\prob\Big\{\dim \; \set{E} \cdot \set{W}_r = tw \Big\}$ in Proposition \ref{prop:dim} and use \eqref{eq:PII}.

\begin{proposition}[Proposition 4 in \cite{BO23}]\label{prop:dim}
	For $\set{W} \in \grassman{w}{q}{m}$ and assuming that $wt < m$, we have 
	\begin{equation*}
		\prob\Big\{\dim\; \set{E} \cdot \set{W}  = tw \;\; \Big \vert \;\; \set{E} \raffect  \grassman{t}{q}{m}\Big\} 
		\; \geq \;   
		1 -  \frac{q^{wt}}{q^m - q^{t-1}}.
	\end{equation*}
\end{proposition}

\end{document}